\documentstyle[preprint,aps,epsf,floats]{revtex}
 
\newcommand{\drawsquare}[2]{\hbox{%
\rule{#2pt}{#1pt}\hskip-#2pt
\rule{#1pt}{#2pt}\hskip-#1pt
\rule[#1pt]{#1pt}{#2pt}}\rule[#1pt]{#2pt}{#2pt}\hskip-#2pt
\rule{#2pt}{#1pt}}

\newcommand{\fund}{\raisebox{-.5pt}{\drawsquare{6.5}{0.4}}}
\newcommand{\Ysymm}{\raisebox{-.5pt}{\drawsquare{6.5}{0.4}}\hskip-0.4pt%
        \raisebox{-.5pt}{\drawsquare{6.5}{0.4}}}
\newcommand{\Yasymm}{\raisebox{-3.5pt}{\drawsquare{6.5}{0.4}}\hskip-6.9pt%
        \raisebox{3pt}{\drawsquare{6.5}{0.4}}}
\newcommand{\antifund}{\overline{\fund}}

\def\NPB#1#2#3{Nucl.~Phys.~B {\bf #1}, #3 (19#2)}
\def\PLB#1#2#3{Phys. Lett. B {\bf #1}, #3 (19#2)}

\def\PRD#1#2#3{Phys. Rev. D {\bf #1}, #3 (19#2)}

\tighten
\draft
\widetext
\preprint{UPR-0991-T, NSF-ITP-02-39, MADPH-02-1268, NI02011-MTH}
\bigskip
\bigskip
\begin{document}
\title{A Three-Family Standard-like Orientifold Model: \\
Yukawa Couplings and Hierarchy}
\medskip
\author{Mirjam Cveti\v c$^{1,2,5}$, Paul Langacker$^{1,2,3,4}$, and
Gary Shiu$^{1,2}$ \vspace{0.2cm}}
\address{$^1$ Department of Physics and
Astronomy, University of Pennsylvania, Philadelphia, PA 19104 \\
$^2$ Institute for Theoretical Physics, University of California,
Santa Barbara, CA 93106 \\
$^3$ School of Natural Sciences, Institute for Advanced Study,
Princeton, NJ 08540 \\
$^4$  Department of Physics, University of Wisconsin, Madison, WI 53706\\
$^5$ Isaac Newton Institute for Mathematical Sciences, University of Cambridge,
Cambridge, U.K.}
\bigskip
\medskip
\maketitle

\def\kl#1{\left(#1\right)}
\def\th#1#2{\vartheta\bigl[{\textstyle{  #1 \atop #2}} \bigr] }

\begin{abstract}
{}We discuss the hierarchy of Yukawa couplings in a supersymmetric three family
 Standard-like string Model. The model is constructed by compactifying
 Type IIA string theory on a $Z_2\times Z_2$
 orientifold in which the Standard Model matter fields arise from intersecting D6-branes.
When lifted to M theory,
 the model amounts to compactification of M-theory on a $G_2$ manifold.
While the actual fermion masses depend on the vacuum expectation values
of the multiple Higgs fields in the model, we calculate the leading
worldsheet
instanton contributions to
the Yukawa couplings and examine the implications of the
Yukawa hierarchy.
\end{abstract}

\section{Introduction}
The basic premise of string phenomenology is to explore the
constructions and  the particle physics implications of
four-dimensional string  solutions  with phenomenologically viable
features  ({\it i.e.}, solutions which
give  rise to an effective theory containing the
Standard Model). The moduli space
of different compactifications of string theory is highly
degenerate at the perturbative level, and so
we are faced with the poorly understood question of
how the string vacuum describing the observable world is
selected. Nevertheless, by
exploring  models with quasi-realistic features from various corners
of M theory, one may deduce some generic
physical implications of string derived models.

Prior to the second string
revolution, the focus of string phenomenology
was on the construction of such solutions
within  the framework of the weakly coupled heterotic string.
Over the years, many
semi-realistic models have been constructed in this framework, and
the resulting phenomenology has been subsequently analysed \cite{review}.
The richness of semi-realistic heterotic string models is also
in sharp contrast to 
the apparent no-go theorem in other formulations of string 
theory \cite{TypeIInogo}.
More recently,
the techniques of conformal field theory in describing D-branes
and orientifold planes allow for the construction of
quasi-realistic string models in another calculable regime of M
theory, as illustrated by the various four-dimensional $N=1$
supersymmetric Type II
orientifolds\cite{ABPSS,berkooz,N1orientifolds,zwart,ShiuTye,KST,afiv,wlm,CPW,kr,aldazabal,CUW}.
In these models, chiral fermions appear on the worldvolume of the
D-branes since they are located at orbifold singularities in the
internal space.

Another promising direction to obtain chiral fermions, which has only recently
been exploited in model building, is to consider branes at angles.
The spectrum of open strings stretched between branes at angles may contain
chiral fermions which are localized at the intersection of branes
\cite{bdl}. This
fact (or its T-dual version, i.e.,
branes with flux)
was employed in
\cite{bgkl,afiru,bkl,Angelantonj,imr,bonn,bklo,bailin,kokor02} in 
constructing semi-realistic brane world models. However,
the semi-realistic models considered in this context are typically 
non-supersymmetric, and the stability of non-supersymmetric
models (and the dynamics involved in restabilization)
is not fully understood. This was one of the motivations of
\cite{CSU1,CSU2,CSU3} in constructing chiral supersymmetric orientifold models
with branes at angles.
The constraints on supersymmetric four-dimensional models are
rather restrictive.
Despite the remarkable progress in developing techniques
of orientifold constructions, 
there is only
one orientifold 
model \cite{CSU1,CSU2,CSU3} that has been constructed so far 
with the ingredients of the MSSM \footnote{Models with features of the
Grand Unified Theories (GUTs) were also constructed in \cite{CSU1,CSU2,CSU3}.}:
${\cal N}=1$ supersymmetry, the
Standard Model  gauge group as a part of the gauge structure, and 
candidate
fields for the three generations of quarks and leptons as well as the
electroweak Higgs doublets.

The general class of supersymmetric orientifold models considered in \cite{CSU1,CSU2,CSU3}
corresponds (in the strong coupling limit) to M theory
compactification on purely geometrical backgrounds admitting a
$G_2$ metric, providing the first explicit  realization  of M
theory compactification on compact $G_2$ holonomy spaces that yields
non-Abelian gauge groups and chiral fermions as well as other quasi-realistic
features of the Standard and GUT models. This work
also sheds light on the recent results
of obtaining
four-dimensional chiral fermions from $G_2$ compactifications of
M theory \cite{AW,Witten,aW,CSU1,CSU2},  as further elaborated in
\cite{CSU3}.

In this paper, we further explore the basic properties of the models,
in particular the three-family Standard-like
Model in \cite{CSU1,CSU2}. The construction, the chiral spectrum
and some of the
basic features of the model were described in the original work
\cite{CSU1,CSU2}.
The details of the chiral and non-chiral
spectra, the  explicit evaluation of the  gauge couplings,
the properties of the two extra $U(1)'$ symmetries, and 
further phenomenological implications associated with charge confinement
in the strongly coupled quasi-hidden sector were discussed in
\cite{CLS}.

The model is not fully realistic.
In addition to the
Standard Model group, there are two
additional $U(1)'$ symmetries,  one of which has family non-universal
and therefore flavor changing couplings,
 and a quasi-hidden non-abelian
sector which becomes strongly coupled above the electroweak scale.
The perturbative spectrum contains a fourth family of exotic ($SU(2)$-
singlet) quarks and leptons, in which, however, the left-chiral states
have unphysical electric charges. In \cite{CLS} it is argued that these
could decouple from the low energy spectrum due to hidden sector charge
confinement, and that anomaly matching requires the physical left-chiral
states to be composites. The model has multiple Higgs doublets and
additional exotic states. The  low energy predictions for the 
gauge couplings depend on the choice moduli parameters. The study  in
\cite{CLS} reveals that  
 $\alpha_{strong}$ can be fitted  to the experimental
value, while $\sin ^2 \theta_W$ and $\alpha_{EM}$ are off by  about a
factor
of $2$ and $3$, respectively.

The purpose of this paper is to carry
out further the
analysis of the couplings in the model. In particular,
we focus on the calculation of Yukawa couplings and
study their physical implications. The Yukawa couplings
among chiral matter are due to the world-sheet instanton
contributions associated with the action of string world-sheet
stretching  among intersections  where
the corresponding chiral matter fields are located. 
 The leading
contribution to the Yukawa couplings is therefore proportional to
$\exp(-A/(2\pi\alpha '))$ where $A$ is the  smallest area  of the
string world-sheet stretching among
the brane intersection points.
The complete calculation of the Yukawa couplings involves
techniques of calculating correlation functions involving
twisted fields in the conformal field theory of open strings.
The origin of the Yukawa couplings, i.e., their world-sheet instanton
origin and the consequences of the exponential hierarchies
within interesecting brane constructions, was first discussed and
analyzed  in
\cite{afiru}.  (For related applications to the fermion mass hierarchy
within GUT intersecting brane constructions,
see \cite{kokor}.)

The purpose of our work is to systematically
evaluate the
leading order contributions to the Yukawa couplings for the
supersymmetric three family Standard-like model. Even though we will
approach the study  only in the leading order of
world-sheet instanton contributions, we shall elucidate these
features explicitly and discuss the consequences of the resulting
hierarchies  of the Yukawa couplings.
The method can also be further applied to other
constructions involving intersecting branes.

The structure of the paper is as follows. In section 2 we
briefly describe the features of the model and the
chiral spectrum. In section 3 we focus on the calculation of the
Yukawa couplings both in the  quark and lepton sectors of the
model. In section 4 we discuss some physical implications of
the hierarchical structure of these couplings and other possible low
energy implications. The conclusions are given in section 5.

\section{Brief Description of the Model}

The model is an orientifold of type IIA on ${\bf T}^6/({\bf
Z}_2\times {\bf Z}_2)$. The orbifold actions have generators
$\theta$, $\omega$ acting as $ \theta: (z_1,z_2,z_3) \to
(-z_1,-z_2,z_3)$, and $\omega: (z_1,z_2,z_3) \to (z_1,-z_2,-z_3)$
on the complex coordinates $z_i$ of ${\bf T}^6$, which is assumed
to be factorizable. The orientifold action  is $\Omega R$, where
$\Omega$ is world-sheet parity, and $R$ acts by $ R:\
(z_1,z_2,z_3) \to ({\overline z}_1,{\overline z}_2,{\overline
z}_3)$. The model contains four kinds of O6-planes, associated with
the actions of $\Omega R$, $\Omega R\theta$, $\Omega R \omega$,
$\Omega R\theta\omega$. The cancellation of the RR crosscap
tadpoles requires an introduction of $K$ stacks of $N_a$
D6-branes ($a=1,\ldots, K$) wrapped on three-cycles (taken to be
the product of 1-cycles $(n_a^i,m_a^i)$ in the $i^{th}$
two-torus), and their images under $\Omega R$, wrapped on cycles
$(n_a^i,-m_a^i)$. In the case where D6-branes are chosen parallel
to the O6-planes, the resulting model is related by T-duality to
the orientifold in \cite{berkooz}, and is non-chiral. Chirality
is however achieved using D6-branes at non-trivial angles.

The cancellation of untwisted tadpoles imposes constraints on the number of
D6-branes and the types of 3-cycles that they wrap around. The
cancellation of twisted tadpoles determines  the orbifold actions on
the Chan-Paton indices of the branes (the explicit form of the
orbifold actions are given in \cite{CSU1,CSU2}).
The condition that the system of branes preserves ${\cal N}=1$
supersymmetry
requires \cite{bdl} that each stack of D6-branes  is related to
the O6-planes by a rotation in $SU(3)$: denoting by $\theta_i$ the angles
the D6-brane forms with the horizontal direction in the $i^{th}$
two-torus, supersymmetry preserving configurations must
satisfy
$
\theta_1\, +\, \theta_2\, +\, \theta_3\, =\, 0
$.
This in turn imposes a constraint on the wrapping numbers and the
complex structure moduli $\chi_i=R_2^{(i)}/R_1^{(i)}$, where
$R_{1,2}^{(i)}$  are the respective sizes of the $i$-th two-torus.

An example leading to a three-family
Standard-like Model
massless spectrum corresponds to the following case. The D6-brane configuration
 is provided in Table \ref{cycles3family},
and satisfies the tadpole cancellation conditions. The
configuration is supersymmetric for $\chi_1:\chi_2:\chi_3=1:3:2$.

\begin{table} \footnotesize
\renewcommand{\arraystretch}{1.25}
\begin{center}
\begin{tabular}{|c||c|l|l|} 
Type & $N_a$ & $(n_a^1,m_a^1) \times
(n_a^2,m_a^2) \times (n_a^3,\widetilde{m}_a^3)$ &  Group \\
\hline
$A_1$ & 8 & $(0,1)\times(0,-1)\times (2,{\widetilde 0})$ &  $Q_{8,8'}$ \\
$A_2$ & 2 & $(1,0) \times(1,0) \times (2,{\widetilde 0})$ &  $Sp(2)_A$ \\
\hline
$B_1$ & 4 & $(1,0) \times (1,-1) \times (1,{\widetilde {3/2}})$ & $SU(2)$  \\
$B_2$ & 2 & $(1,0) \times (0,1) \times (0,{\widetilde {-1}})$ & $Sp(2)_B$ \\
\hline
$C_1$ & 6+2 & $(1,-1) \times (1,0) \times (1,{\widetilde{1/2}})$ &
 $ SU(3), Q_3,Q_1 $ \\
$C_2$ & 4 & $(0,1) \times (1,0) \times (0,{\widetilde{-1}})$  & $Sp(4)$  \\
\end{tabular}
\end{center}
\caption{\small D6-brane configuration for the three-family model. Here,
$\tilde{m}^3_a = m_a^3 + \frac{1}{2} n_a^3$.}
\label{cycles3family}
\end{table}

The rules to compute the spectrum are analogous to those in
\cite{bkl}. Here, we summarize the resulting chiral spectrum in
Table \ref{matter}, found in \cite{CSU1,CSU2}, where
\begin{equation} I_{ab}\ =\
(n_a^1m_b^1-m_a^1n_b^1)(n_a^2m_b^2-m_a^2n_b^2)(n_a^3m_b^3-m_a^3n_b^3)
\label{internumber}
\end{equation}
is the intersection number of $D6_a$ and $D6_b$ branes \cite{bgkl,afiru}.

\medskip
\begin{table} \footnotesize
\renewcommand{\arraystretch}{1.25}
\begin{center}
\begin{tabular}{|c|c|}
{\bf Sector}   &
{\bf Representation} \hspace{4cm} \\
\hline\hline
$aa$    &  \hspace{1cm} $U(N_a/2)$ vector multiplet \hspace{4cm} \\
       & \hspace{1cm} 3 Adj. chiral multiplets  \hspace{4cm} \\
\hline\hline
$ab+ba$   &  $I_{ab}$ chiral multiplets in
$(\fund_a,\antifund_b)$ rep.  \hspace{4cm} \\
\hline\hline
$ab'+b'a$ &  $I_{ab'}$ chiral multiplets in $(\fund_a,\fund_b)$ rep.
  \hspace{4cm} \\
\hline\hline
$aa'+a'a$ &  $-\frac 12 (I_{aa'} - \frac{4}{2^{k}} I_{a,O6})$
chiral multiplets in $\Ysymm$ rep. \hspace{4cm}  \\
          & $-\frac 12 (I_{aa'} + \frac{4}{2^{k}} I_{a,O6})$
chiral multiplets in $\Yasymm$ rep.  \hspace{4cm} \\
\end{tabular}
\end{center}
\caption{\small General spectrum on D6-branes at generic angles
(namely, not parallel to any O6-plane in all three tori). The spectrum is
valid for tilted tori. The models may contain additional non-chiral pieces
in the $aa'$ sector and in $ab$, $ab'$ sectors with zero intersection, if
the relevant branes overlap.
\label{matter} }
\end{table}

\begin{table} \footnotesize
\renewcommand{\arraystretch}{1.25}
\begin{center}
\begin{tabular}{|c||c||c|c|c|c|c||c|c|c|}
Sector & $SU(3)\times SU(2)\times Sp(2)_B\times Sp(2)_A\times Sp(4)$ &
$Q_3$ & $Q_1$ & $Q_2$ & $Q_8$ & $Q_8'$ & $Q_Y$ & $Q_8-Q_8'$ & Field
\\
\hline
$A_1 B_1$ & $3 \times 2\times (1,{\overline 2},1,1,1)$ &
0 & 0 & $-1$ & $\pm 1$ & 0 & $\pm \frac 12$ & $\pm 1$ & $H_U$, $H_D$\\
          & $3\times 2\times (1,{\overline 2},1,1,1)$ &
0 & 0 & $-1$ & 0 & $\pm 1$ & $\pm \frac 12$ & $\mp 1$ & $H_U$, $H_D$\\
$A_1 C_1$ & $2 \times (\overline{3},1,1,1,1)$ &
$-1$ & 0 & 0 & $\pm 1$ & 0 & $\frac 13, -\frac 23$ & $1,-1$ &
$\bar{D}$, $\bar{U}$\\
          & $2 \times (\overline{3},1,1,1,1)$ &
$-1$ & 0 & 0 & 0 & $\pm 1$ & $\frac 13, -\frac 23$ & $-1,1$ &
$\bar{D}$, $\bar{U}$\\
          & $2 \times (1,1,1,1,1)$ &
0 & $-1$ & 0 & $\pm 1$ & 0 & $1,0$ & $1,-1$ &
$\bar{E}$, $\bar{N}$\\
          & $2 \times (1,1,1,1,1)$ &
0 & $-1$ & 0 & 0 & $\pm 1$ & $1,0$ & $-1,1$ &
$\bar{E}$, $\bar{N}$\\
$B_1 C_1$ & $(3,{\overline 2},1,1,1)$ &
1 & 0 & $-1$ & 0 & 0 & $\frac 16$ & 0 & $Q_L$\\
             & $(1,{\overline 2},1,1,1)$ &
0 & 1 & $-1$ & 0 & 0 & $-\frac 12$ & 0 & $L$\\
$B_1 C_2$ & $(1,2,1,1,4)$ &
0 & 0 & $1$ & 0 & 0 & 0 & 0 & \\
$B_2 C_1$ & $(3,1,2,1,1)$ &
1 & 0 & 0 & 0 & 0 & $\frac 16$ & 0 & \\
          & $(1,1,2,1,1)$ &
0 & 1 & 0 & 0 & 0 & $-\frac 12$ & 0 & \\
$B_1 C_1^{\prime}$ & $2\times (3,2,1,1,1)$ &
1 & 0 & 1 & 0 & 0 & $\frac 16$ & 0 & $Q_L$ \\
                   & $2\times (1,2,1,1,1)$ &
0 & 1 & 1 & 0 & 0 & $-\frac 12$ & 0 & $L$ \\
\hline
$B_1 B_1^{\prime}$ & $2\times (1,1,1,1,1)$ &
0 & 0 & $-2$ & 0 & 0 & 0 & 0 &  \\
                   & $2\times (1,3,1,1,1)$ &
0 & 0 & $2$ & 0 & 0 & 0 & 0 & \\
\hline \hline 
$A_1 A_1$ & $3 \times 8 \times (1,1,1,1,1)$ & 0 & 0 & 0 & 0 & 0 & 0 & 0 &  \\
& $3 \times 4 \times (1,1,1,1,1)$ & 0 & 0 & 0 & $\pm 1$ & $\pm 1$ & $\pm 1$
& 0 &  \\
& $3 \times 4 \times (1,1,1,1,1)$ & 0 & 0 & 0 & $\pm 1$ & $\mp 1$ & 0
& $\pm 2$ &  \\
& $3 \times  (1,1,1,1,1)$ & 0 & 0 & 0 & $\pm 2$ & 0 & $\pm 1$ & $\pm 2$ &  \\
& $3 \times  (1,1,1,1,1)$ & 0 & 0 & 0 & 0 & $\pm 2$ & $\pm 1$ & $\mp 2$ &  \\
$A_2 A_2$ & $3 \times (1,1,1,1,1)$ & 0 & 0 & 0 & 0 & 0 & 0 & 0 &  \\
$B_1 B_1$ & $3 \times (1,3,1,1,1)$ & 0 & 0 & 0 & 0 & 0 & 0 & 0 &  \\
 & $3 \times (1,1,1,1,1)$ & 0 & 0 & 0 & 0 & 0 & 0 & 0 &  \\
$B_2 B_2$ & $3 \times (1,1,1,1,1)$ & 0 & 0 & 0 & 0 & 0 & 0 & 0 &  \\
$C_1 C_1$ & $3 \times (8,1,1,1,1)$ & 0 & 0 & 0 & 0 & 0 & $0$ & 0 &  \\
& $3 \times (1,1,1,1,1)$ & 0 & 0 & 0 & 0 & 0 & $0$ & 0 &  \\
$C_2 C_2$ & $3 \times (1,1,1,1,5+1)$ & 0 & 0 & 0 & 0 & 0 & 0 & 0 &  \\
\end{tabular}
\end{center}
\caption{\small The chiral spectrum of the open string sector in the
three-family model. 
To be complete, we also list  (in the bottom part of the table,
below the double horizontal line) the chiral states from the
$aa$ sectors, which are not localized at the intersections.
\label{spectrum3}}
\end{table}

The chiral spectrum is given in Table \ref{spectrum3} (see \cite{CSU1}).
Here, we list also the chiral matter from the $aa$ sectors.
The charges of the matter fields under various $U(1)$ gauge fields
of the model are tabulated.
The generators
$Q_3$, $Q_1$ and $Q_2$ refer to the $U(1)$ factor within the corresponding
$U(n)$, while $Q_8$, $Q_8^{\prime}$ are the $U(1)$'s arising from
Higgsing the $USp(8)$.
The last column provides the
charges under a particular anomaly-free $U(1)$ gauge field:
\begin{equation}
Q_Y = \frac{1}{6} Q_3 - \frac{1}{2} Q_1 + \frac{1}{2} \left( Q_8 + Q_8^{\prime}
\right)
\end{equation}
This linear combination $Q_Y$
plays the role of hypercharge. 
There are two additional non-anomalous $U(1)$ symmetries, i.e., 
$\frac{1}{3} Q_3 - Q_1$ and $Q_8 - Q_{8}^{\prime}$.
The spectrum of chiral
multiplets corresponds to three
quark-lepton generations, a number of vector-like Higgs
doublets, and an anomaly-free set of chiral matter.
It includes states corresponding
to the right-handed $SU(2)$-singlet fields of a fourth
family. However, their natural left-handed partners, from
the $B_2C_1$ sector, have the wrong hypercharge. It is argued in~\cite{CLS}
that these disappear from the low energy spectrum
due to the strong coupling of the first $Sp(2)$ group, to be
replaced by composites with the appropriate quantum numbers
to be the partners of the extra family of right-handed fields.

The gauge couplings  of the various gauge fields in the model
(determined by  the
volume of the 3-cycles  that  the corresponding  D6-brane  wraps
around \cite{CSU1}) were calculated explicitly in \cite{CLS}. In this
paper we shall focus on the Yukawa couplings.

\section{Yukawa Couplings}
\label{yukawasection}

In this section we calculate the leading contribution to
the Yukawa couplings among the chiral matter fields in the model.
The  string theory  calculation of the Yukawa couplings requires the
techniques of computing string amplitudes that involve
twisted fields of the conformal field theory describing the open strings
states at each intersection.  
In particular, the quantization of the open string sector associated with
the string states at the interection of two  D-branes  at a general angle
$\theta$ involves states with  the boundary conditions 
that are a  linear combination of Dirichlet and Neumann
boundary conditions. Thus the mode expansion is in terms of
$\alpha_{n-k}$ modes  and the  non-integer powers of
the world-sheet coordinate $z\equiv {\rm exp}(\tau+i\sigma)$, i.e.,
$X^i\sim
\sum_{n}{{\alpha_{n-k}^i}\over {n-k}}\, z^{-(n-k)}$,  where $n$ are
integers, and
$k= {\theta\over{2\pi}}$. Therefore   states in this sector  are created
by acting with 
 $\alpha_{n-k}$ ($n-k< 0$) on 
the ``twisted'' vacuum $\sigma_{k}|\, 0 \, \rangle$, where $\sigma_{k}$ is the
conformal field that ensures the correct boundary conditions on the open
string states.   As a consequence the string amplitude for  the three
states, each of them at an intersection of two  D-branes
where the three intersections form the edges of a triangle, involves a
calculation of a correlator of the type $\langle \, 0\, | \partial_z
X^{i_1}\, 
\partial_z X^{i_2}\, 
\partial_z X^{i_3}\, 
\sigma_{k_1}\, \sigma_{k_2}\, \sigma_{k_3}\,| \, 0\,  \rangle $ (with
$\sum_{i=1}^3k_i=1$). [The
fermionic sector of the correlator can be determined in a
straightforward way by employing the
bosonisation procedure of the world-sheet fermionic degrees of
freedom.]  Since each state is localized at the intersection of
the D-branes, this  amplitude involves 
the contribution of the worldsheet instantons, and it is thus
exponentially suppressed by the area of the
corresponding intersection  triangle. The results of the
calculation  should be analogous  to Yukawa coupling calculations  for the
twisted
closed string states of orbifolds~\cite{Dixonetal}. However, the
subtleties   of the open-string sector calculations (such as
the so-called ``doubling trick'', that allows one to express the open
string modes in terms of the holomorphic world-sheet coordinate $z$, only;
$z$  is now defined on the whole complex plane, along with the
boundary conditions  for states specified on the real line)   require
further study.

The  Yukawa  couplings can therefore be expressed as  a  sum over  the
worldsheet
instantons associated with the action of the string worldsheet
stretching  among the intersection points  where the
corresponding  chiral matter fields are located. The couplings  
are schematically of the form: $\sum_{n=1}^\infty
Z_n
\exp[-(n c_n A)/(2\pi \alpha')]$. Here  $A$ is the smallest
area  of the triangle associated with the corresponding brane
intersections  and $\alpha^{\prime}$ is the string tension, related
to the string scale by $M_s = ( \alpha^{\prime})^{-1/2}$. (The
factor  $1/(2\pi)$ in the exponents is due to  the normalization of the
string Nambu-Goto action with the pre-factor $1/(2 \pi \alpha ')$.) The
pre-factors   $Z_n$
and     the coefficients $c_n$ in the exponents are  of  ${\cal{O}} (1)$.
(The coefficients $c_n$ should in principle include the multiplicity
factors due to the orbifold and orientifold symmetries.)
The
leading contribution to the Yukawa couplings is therefore
proportional to $Z_1\exp[-(c_1A)/(2 \pi \alpha ')]$.
The  world-sheet instanton
origin  of Yukawa couplings and the implications for hierarchies 
within interesecting D-branes was originally  studied in
\cite{afiru}.

At this  stage we shall
approach the study systematically  by studying the leading order
contributions, only. Within this context we shall
evaluate the intersection  areas $A$  explicitly in terms of the
moduli of internal tori. Indeed, even in the leading order
in the determination of the Yukawa couplings there remains an
uncertainty, since $Z_1$ and $c_1$,  which are coefficients
of ${\cal O}(1)$, 
can only be determined by an explicit string calculation.
(Note also  that 
the physical values of the  Yukawa couplings also depend on the normalization
of the kinetic energy terms for the corresponding matter field,
which we will not address here either.)

In particular, we shall  explore the basic building blocks for
the calculation of $A$, by first positioning the branes  very close to the
symmetric positions in the six-torus.
As the next step we shall then
explore the consequences for the Yukawa coupling  hierarchy
when the
branes are moved from the symmetric positions.

There are couplings between the $A_1 B_1$, $B_1 C_1'$
and $C_1' A_1$ sectors. Since the $C_1' A_1$ sector is the same as $A_1' C_1$,
 and furthermore, $A_1=A_1'$ (because the $A_1$ brane is the same
as its own orientifold image),  in principle
there are non-zero couplings of the form $(A_1 B_1)(B_1 C_1')(A_1 C_1)$,
which could give rise to the Yukawa couplings of the two families of quarks 
and leptons from the $B_1 C_1'$ sector (see Table \ref{spectrum3}).
The third family has {\it no}  Yukawa couplings, since
the left-handed quarks and leptons in this family arise from 
the $B_1 C_1$ sector instead, and hence the three-point couplings
are not gauge invariant.

\begin{center}
\begin{figure}[ht]
\epsfysize=7cm
\epsfbox{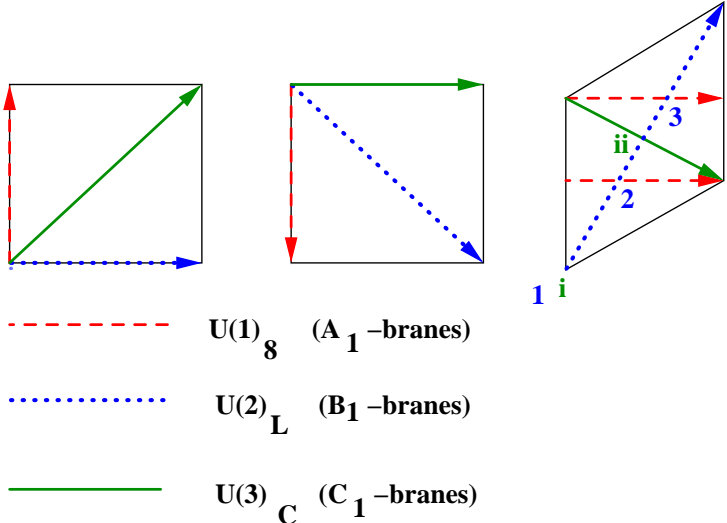} 
\caption{The initial symmetric
configuration of the $A, B, C'$ sectors of branes, associated with
the $U(1)_{8,8'}$ , $U(2)_L$ and $\{U(3)_C, U(1)_1 \}$
sectors, are denoted by dashed, dotted,
and solid lines, respectively.  The intersections denoted by
$\alpha=(1,2,3)$ and $\gamma=(i,ii)$ correspond to the appearance of
Higgs and left-handed families, respectively. }
\label{basic}
\end{figure}
\end{center}

The basic ingredients for calculating the intersection  areas 
are given in Figure \ref{basic}.

This is an initial symmetric configuration of the $A, B, C'$
sectors of branes, associated with the $U(1)_{8,8'}$ , $U(2)_L$
and $\{U(3)_C, U(1)_1 \}$ sectors, respectively. The set of
$A$ branes, associated with $U(1)_8$ and $U(1)_{8'}$, are
positioned  very close  to the corresponding orientifold plane.
(Had they all been positioned exactly on top of the  orientifold
plane, the gauge group would have been enhanced to $USp(8)$).
Thus the couplings associated with the pairs of states that are
charged under $U(1)_8$ and $U(1)_{8'}$, respectively are
approximately {\it degenerate}. We denote the two sets of Higgs
fields with $U(1)_8$ charges 
 as $H^{\alpha}_{\{U,D\},\{I,II\}}$ where $\alpha
=\{1,2,3\}$.
Here, $\alpha$ labels the intersection points of the
$A_1$ and $B_1$ branes (where the Higgs fields are located).
The pairs of states denoted by
$\{I,II\}$ indices correspond to the two sets of fields
appearing at the same intersections.
Analogous notation is used for the corresponding right-handed
quark and lepton sector. The set of fields associated with
$U(1)_{8'}$ charges are denoted by $H\to H'$ and $\{\bar{U},\bar{D}\}\to
\{\bar{U}',\bar{D}'\}$.

 We have
also positioned branes associated with $U(3)_C$ and $U(1)_1$
nearby, which ensures at this stage the near  degeneracy of the
couplings associated with the  quark doublets $Q_L^{\gamma}$ and leptons
$L_L^{\gamma}$ ($\gamma=(i,ii)$) as well as that of the
Up- and Down-sector.  Due to this large degeneracy, we shall
only describe the couplings for the Up-quark sector.

 From Figure \ref{basic}, which depicts the location of the intersections of
the $A,
B,C'$ branes,
  it is evident that there are  different
 Yukawa couplings associated with  the location of the intersections of the
 two types of
 left-handed quarks $\gamma =(i,ii)$ and the location of the three
 types of
 Higgs fields $H_{U\, \{I,II\}}^{\alpha}$ where
$\alpha =1,2,3$ (and $H\to H'$ sectors).

While the Higgs fields  (and the right-handed quarks)
associated with index $I$ and $II$ formally appear at the same
intersection, the orientifold and orbifold projection in the
construction of  these states ensure that  only  pairs of the
Higgs and right-handed quarks  with the same $I$ or $II$ index
couple to each other.

Thus, in this degenerate case, the Yukawa interactions
take the form:
\begin{equation}
H_{yukawa}= \sum_{\alpha, \gamma}h_{\alpha,\, \gamma}Q_L^{\gamma}
(\bar{U}_IH^{\alpha}_{U\, I}+{\bar{U}_{II}} { H}^{\alpha }_{U\, II})\, +
(\{\bar{U},H_U\}\to \{\bar{U}',H_U'\}) , \label{Yukawa}
\end{equation}
where $h_{\alpha, \gamma}\sim {\rm exp}(-A_{\alpha
,\gamma}/(2 \pi\alpha'))$.

The area of the  triangle associated with the three intersection points
(in the six dimensional internal space)  can be
calculated in terms of the  products of the vectors ${\vec a}$,
${\vec b}$ and ${\vec c}$, specifying  the respective locations
of the three intersections points:
\begin{equation}
Area= \textstyle{1\over 2}|[{\vec a}-{\vec b}]\times  [{\vec a}-{\vec c}]|=
\textstyle{1\over 2}\sqrt{[{\vec a}-{\vec b}]^2 [{\vec a}-{\vec c}]^2-\left([{\vec
a}-{\vec b}] \cdot [{\vec a}-{\vec c}]\right)^2} \, . \label{area}
\end{equation}

\begin{figure}[t]
\centering
\begin{minipage}[c]{0.5\textwidth}
\centering
 \epsfysize=7cm\epsfbox{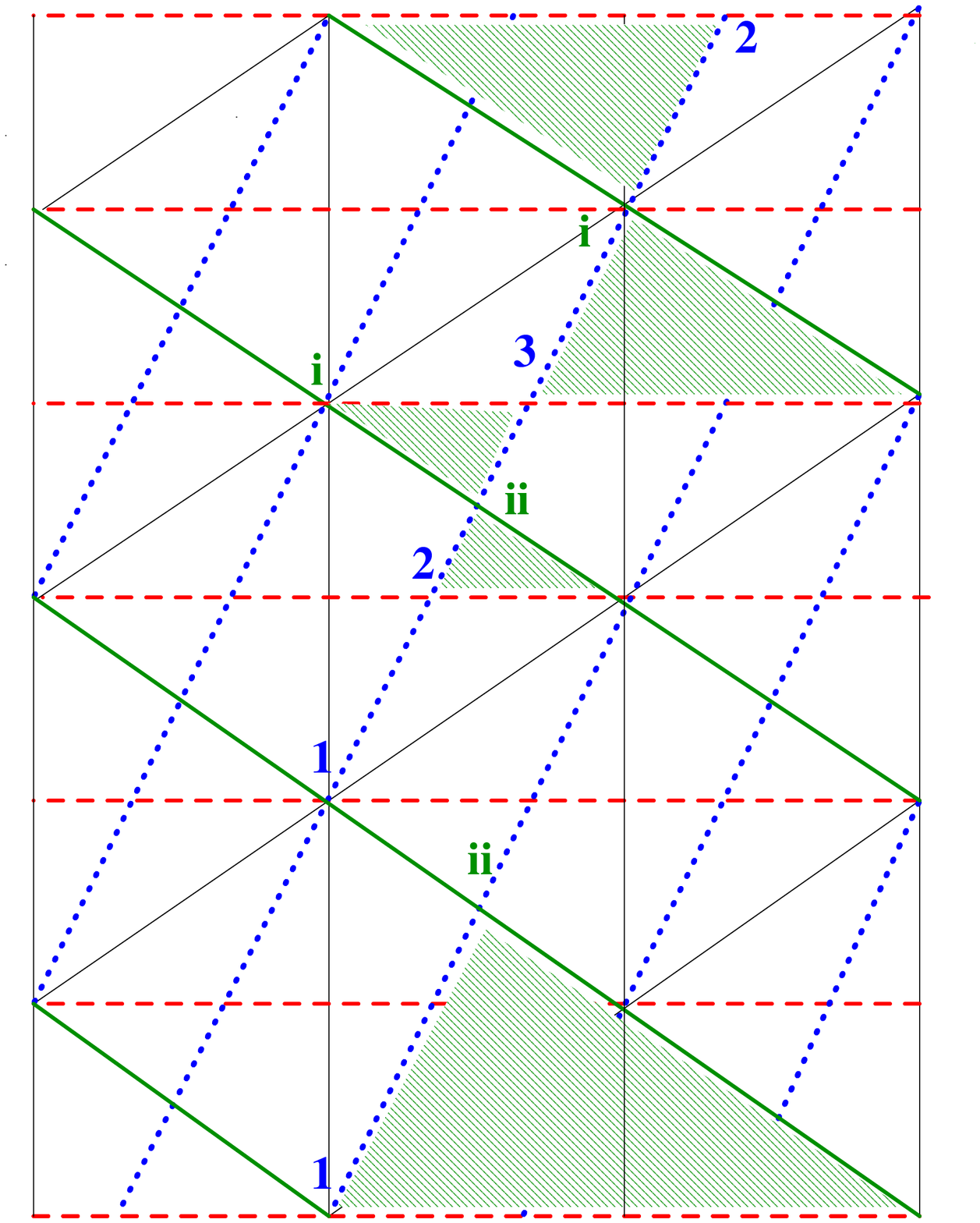}
 \end{minipage}%
\begin{minipage}[c]{0.5\textwidth}
\centering
 \caption{Intersection
areas of the branes in the third torus. The thin solid lines
denote the lattice, the thick solid lines-the $U(3)_C$ branes, the
dotted lines-$U(2)_L$ branes and the  dashed ones- $U(1)_{8,8'}$
branes. Again $\alpha=(1,2,3)$ and $\gamma=(i,ii)$ denote the
location of the three Higgs fields and the two left-handed quark
families, respectively. }
\label{yuk3}
\end{minipage}
\end{figure}

After these preliminaries we are set to calculate the minimal
intersection areas $A_{\alpha, \gamma}$. These can be easily
determined from Figure \ref{basic}, which depicts the position of the
building block branes in the fundamental domain of the toroidal
lattice, and Figure \ref{yuk3} which depicts the    relevant
intersection areas for  the third toroidal lattice.

 Employing eq.
(\ref{area}) we obtain the straightforward results for the
intersection areas:
\begin{eqnarray}
A_{1,\,i}&=&0\nonumber\\
A_{2,\, i}&=& A_{3,\, i}=\textstyle{1\over {3}} R_{1}^{(3)}
R_2^{(3)}\nonumber\\
A_{2,\, ii}&=& A_{3,\, ii}=\textstyle{1\over
{12}} R_{1}^{(3)}
R_2^{(3)}\nonumber\\
A_{1,\, ii}&=& \textstyle{3\over {4}} R_{1}^{(3)}
R_2^{(3)}\label{area0}
\end{eqnarray}
where $R^{(i)}_{1,2}$ refer to the two sizes of the $i$-th torus,
along the $x_i$ and $y_i$
axis respectively\footnote{This notation differs slightly from~\cite{CLS},
in which $R_{1,2}^{(i)}$ represented radii, i.e., $R_{1,2}^{(i)}$ in this 
paper corresponds to $2\pi\, R_{1,2}^{i}$ in \cite{CLS}.}.
Note also that $R^{(i)}_1 R^{(i)}_2$ 
corresponds to the
area of the i-th two-torus.
Due to the symmetry of the  configuration there
is no contribution from the area arising from the first two
two-tori.
 We can therefore encounter a sizable hierarchy among different Yukawa
couplings. In particular the sub-leading terms
$h_{1,ii}$ for $\alpha=1$ are smaller than the couplings
for $\alpha=2,3$, as can be seen in~(\ref{area0}).

There are phenomenological constraints on the possible
values of $R_{1,2}^{(i)}$. 
The Planck scale and various Yang-Mills couplings are related to the string
coupling $g_s$ by
\begin{equation}
(M_P^{(4d)})^2 = \frac{ M_s^8 V_6 }{ (2 \pi)^7 g_s^2}\, ,
\label{mpms}
\end{equation}
and
\begin{equation}
\frac{1}{g_{YM}^2} = \frac{ M_s^3 V_3}{(2 \pi)^4 g_s}\, ,
\label{gaugec}\end{equation}
where $V_6$ is the volume of the six-dimensional orbifold and  $V_3$ is
the volume of the three-cycle that a specific set of D6-brane wraps. (These
volume factors  have
been explicitly calculated in \cite{CLS} in terms of the wrapping numbers
$(n^i,m^i)$ and  $R_{1,2}^{(i)}$.) 

Using (\ref{mpms}) and (\ref{gaugec}) one can
eliminate $g_s$ and obtain the relationship  between $g_{YM}$, $M_P^{4d}$
and $M_s$:
\begin{equation}
g_{YM}^2 M_P^{(4d)} = \sqrt{2 \pi} M_s \frac{\sqrt{V_6}}{V_3}~\, .
\label{gaugep}
\end{equation}
For a fixed  value of 
$M_s/M_P^{(4d)}$, the  $g_{YM}$ depend only on the ratios
$\frac{\sqrt{V_6}}{V_3}$, which are functions of 
 the complex structure moduli
$\chi_i=R^{(i)}_2/R^{(i)}_1$ only,  and have been explicitly evaluated in
\cite{CLS}.

Since each gauge group factor of
the Standard Model arises from a separate set of branes wrapping
a specific three-cycle, there is no internal direction transverse to all
the branes. It therefore follows from
(\ref{mpms},\ref{gaugec},\ref{gaugep})
that the large Planck scale $M_P^{4d}$ cannot be generated by taking any
of the internal directions much larger than the inverse of the string
scale $M_s$, since for  perturbative values of the string coupling $g_s$
that would
make (at least one of) the gauge couplings unrealistic. 
Thus a large Planck scale is  generated from a large string
scale   and not from a large volume,  
which is  then also compatible with the gauge coupling constraints
(\ref{gaugec},\ref{gaugep}).
(Note also  that experimental bounds on the Kaluza-Klein modes
of the Standard Model gauge bosons
imply that the extra dimensions cannot be larger than ${\cal
O}({\rm TeV}^{-1})$, but  this is a much weaker bound than the one  obtained
by the arguments above.) 
Finally, the $R_{1,2}^{(i)}$'s cannot be much smaller
than the string scale $M_s^{-1}$ as this would again make 
the Planck scale and gauge couplings unrealistic.
One should however point out that there  still remains some  flexibility
in adjusting the sizes $R_{1,2}^{(i)}$'s by an order of magnitude or so away
from ${\cal O}(M_s^{-1})$. 

The above  constraints that limit generic values of the sizes
$R_{1,2}^{(i)}$'s to be close to the inverse of the string scale
$M_s^{-1}$  (and Planck scale $M_P^{4d}$ close to $M_s$) 
have implications for the hierarchy of
the  Yukawa couplings.  Had  one had $R_{1,2}^{(i)}\gg M_s^{-1}$
the couplings would have been exponentially suppressed.
However, since $R_{1,2}^{(i)}={\cal O}(M_s^{-1})$, the range dictated from
constraints on the Planck scale and gauge couplings, the hierarchy
among Yukawa couplings is non-degenerate  and may potentially have
interesting phenomenological implications.
 For definiteness, we will require $M_s R_{1,2}^{(i)} \ge 2 \pi$.
\begin{center}
\begin{figure}[ht]
\epsfysize=5cm
\epsfbox{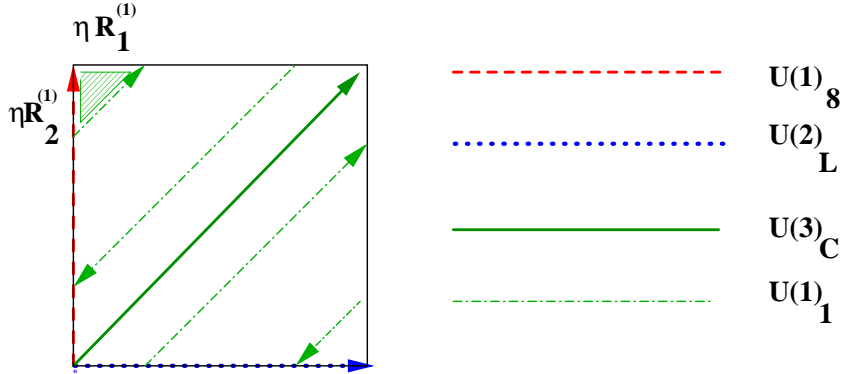} 
\bigskip
\caption{The brane configurations in the first torus, depicting the
breaking on $U(4)$ Pati-Salam symmetry down to $U(3)_C\times U(1)_1$.
The $U(1)_1$
branes (denoted by dash-dotted line) are positioned in a $Z_2$ symmetric
way relative to $U(3)_C$ branes (denoted by a solid line). The
separation between them  is $\eta R_{1,2}^{(1)}$ in the respective $x$- and
$y$-directions. The relevant interesection area in the 
first torus, contributing to the lepton Yukawa coupling is denoted 
by a shaded area. [$U(2)_L$
and $U(1)_{8,8'}$ branes are denoted by a dotted and a dashed line,
respectively.]
}\label{leptonF}
\end{figure}
\end{center}

\subsection{Lepton-Quark Splitting}

The  eight $C_1$-branes are split in sets of six and two, thus
ensuring the  breakdown of  $U(4)$ (Pati-Salam type) symmetry down to
 $U(3)_C$ and $U(1)_1$.  We chose  to split them in the first two-torus,
 keeping $U(3)_C$ along the $Z_2$ symmetric position and moving $U(1)_1$
 branes  relative to $U(3)_C$ ones by
a distance $\eta R^{(1)}_{1,2}$ away in the
 $x$- and $y$- direction, respectively. (See Figure \ref{leptonF}).
 It now becomes a straightforward exercise to determine the
new areas associated
 with the lepton Yukawa couplings. 
The  areas associated with the lepton Yukawa couplings can be expressed in terms of  the
 areas for the quark Yukawa couplings by

\begin{equation}
A^{lepton}_{\alpha, \, \gamma}=\textstyle{1\over 2}\sqrt{
(\eta\,{R^{(1)}_1})^2\,(\eta\, {R^{(1)}_2})^2 +
(\eta\, {R^{(1)}_1})^2\, {\vec {\cal A}_{\alpha,\,\gamma}}^2+
(\eta{R^{(1)}_2})^2\, {\vec {\cal B}_{\alpha,\,\gamma}}^2
+4{A^{quark}_{\alpha,\,\gamma}}^2},\label{arealep}
\end{equation}
where  ${\vec {\cal A}_{\alpha,\,\gamma}}$ and ${\vec {\cal
B}_{\alpha,\,\gamma}}$ specify the vectors 
for the  respective $U(1)_{8,8'}$ and $U(2)_L$   sides  of the
triangles for
the corresponding  $(\alpha,\, \gamma)$  intersections (in the third 
toroidal direction).

The areas (\ref{arealep}) for lepton Yukawa couplings are always larger than
  those of the quark couplings.
   This formula
   is  valid as long as  $\eta$ is less or $\sim 1/2$.
The values of ${\vec {\cal A}_{\alpha,\,\gamma}}$ and ${\vec {\cal
B}_{\alpha,\,\gamma}}$ are give in Table~\ref{veclep},
while $A^{quark}_{\alpha,\,\gamma}$ are listed in eq.~(\ref{area0}).

\begin{table} 
[htb] \footnotesize
\renewcommand{\arraystretch}{1.25}
\begin{center}
\begin{tabular}{|c|c|c|}
$(\alpha,\gamma)$   & ${\vec {\cal A}_{\alpha,\,\gamma}}^2$ \hspace{2cm}
 & ${\vec {\cal B}_{\alpha,\,\gamma}}^2$ \hspace{2cm} \\
\hline
$(1,i)$ & 0 \hspace{2cm} & 0 \hspace{2cm} \\
\hline
$(2,i),(3,i)$    & $\left(\textstyle{4\over 3}R^{(3)}_1\right)^2$ \hspace{2cm}
& $\left(\textstyle{1\over 3}R^{(3)}_1\right)^2+\left(\textstyle{1\over
2}R^{(3)}_2\right)^2$ \hspace{2cm} \\
\hline
$(2,ii),(3,ii)$    & $\left(\textstyle{2\over 3}R^{(3)}_1\right)^2$
\hspace{2cm}
& $\left(\textstyle{1\over 6}R^{(3)}_1\right)^2+\left(\textstyle{1\over
4}R^{(3)}_2\right)^2$ \hspace{2cm} \\
\hline
$(1,ii)$           & $\left(2R^{(3)}_1\right)^2$ \hspace{2cm}
& $\left(\textstyle{1\over 2}R^{(3)}_1\right)^2+\left(\textstyle{3\over
4}R^{(3)}_2\right)^2$ \hspace{2cm}
\end{tabular}
\end{center}
\caption{The values of ${\vec {\cal A}_{\alpha,\,\gamma}}$ and ${\vec {\cal
B}_{\alpha,\,\gamma}}$ for the  respective $U(1)_{8,8'}$ and
$U(2)_L$  sides 
of the
intersection triangles in the third torus for various $\alpha$ and
$\gamma$.} \label{veclep}
\end{table}


\subsection{Up-Down Yukawa Coupling Splitting}

\begin{figure}[ht]
\centering
\epsfysize=7cm\epsfbox{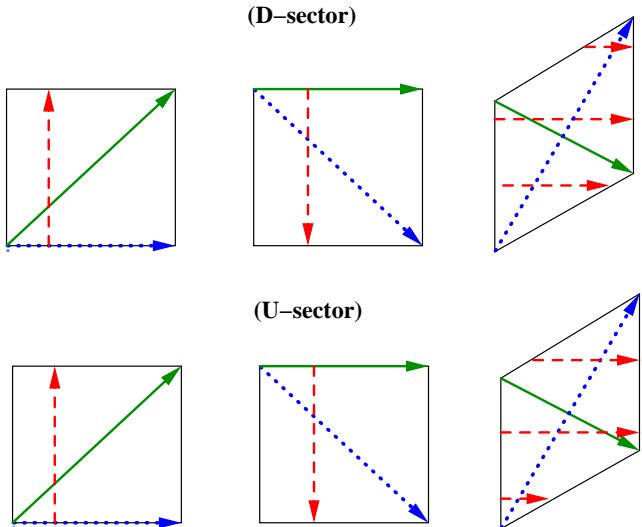}
\caption{The splitting of A-type
branes (associated with $U(1)_{8,8'}$ and denoted by the dashed
lines) from the orientifold planes. For simplicity the figure
shows only the fundamental domain of each of the three two-tori.
The solid and dotted lines denote the $U(3)_C$ and $U(2)_L$
branes, respectively.}
\label{yuk2}
\end{figure}

The degeneracy  of the Yukawa couplings that are associated with
states charged under $U(1)_8$ and $U(1)_{8'}$ can be removed by
splitting the branes associated with  the first and second
abelian factors from the orientifold plane by a distance $\epsilon_i$
and $\epsilon_i'$  in the $i$-th torus ($i=1,2,3$) (see Figure \ref{yuk2}).
 This in turn provides a mechanism for Up-Down sector splitting.

\begin{figure}[ht]
 \epsfxsize=8cm \epsfbox{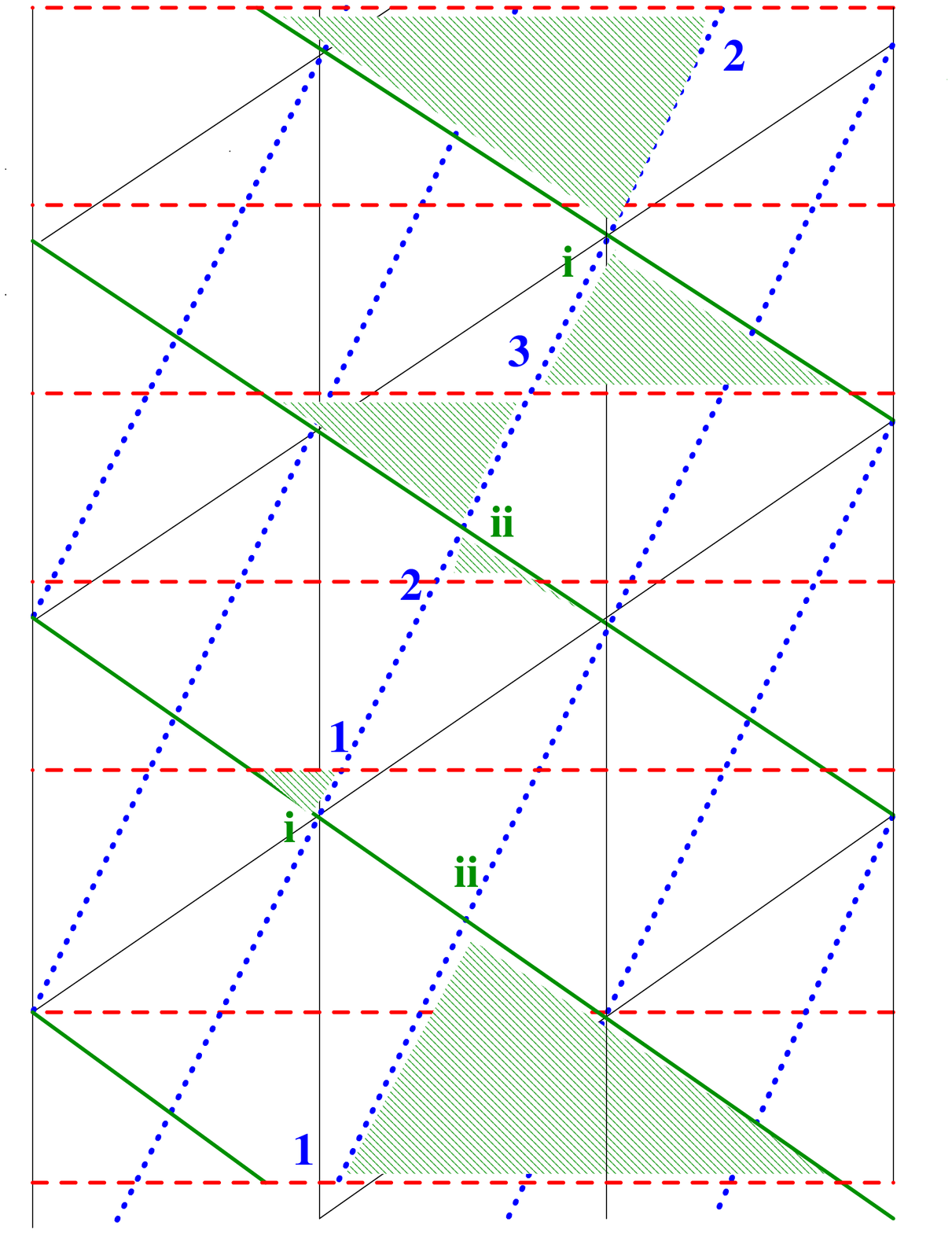}
 \caption{Relevant intersection areas in the third toroidal lattice for the
 Up-sector.}\label{yuk4}
 \end{figure}

\begin{figure}[ht]
 \epsfxsize=8cm\epsfbox{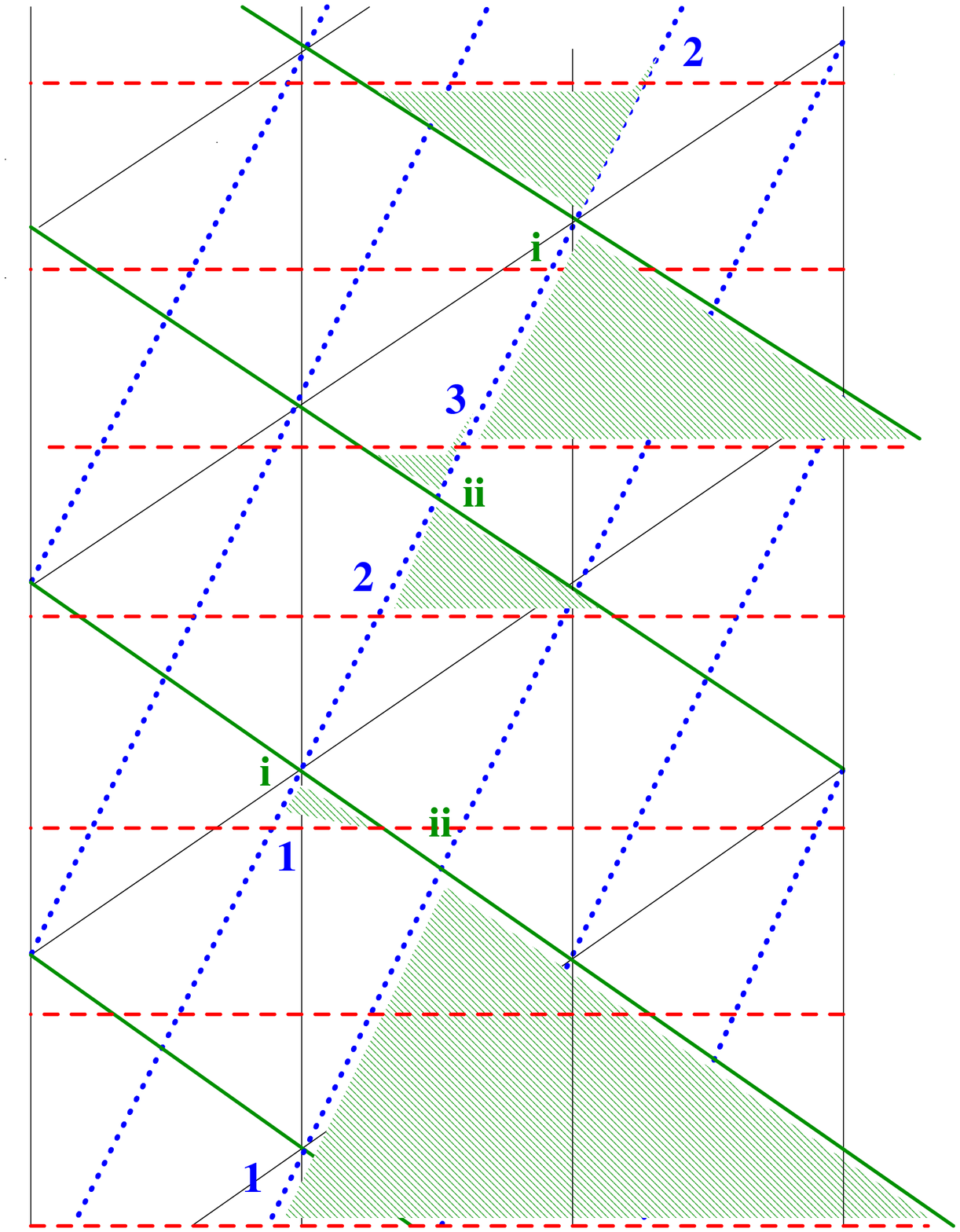}
 \caption{Relevant intersection areas in the third toroidal lattice for
the
 Down-sector.}\label{yuk5}
 \end{figure}

 One
can show that the basic ingredients for determining the Up-type
[Down-type] Yukawa couplings is to study the
intersection of the A-type branes (associated with $U(1)_{8,8'}$)
  moved by a
distance $+\epsilon_{i}R^{(i)}_1$ [$+\epsilon_{i}R^{(i)}_1$]
away from the  (vertical) orientifold planes
 in the first two ($i=(1,2)$)
two-tori, and a distance  $+\epsilon_3\, R^{(3)}_2$
[$-\epsilon_3\, R^{(3)}_2$ ] from the (horizontal) orientifold
plane in the third torus.
 Figure
\ref{yuk2} depicts these basic displacements of the $A$-type branes
 in the
fundamental domain of each of the two-tori for the Up- and
Down-sectors, respectively. In addition, Figures \ref{yuk4} and
\ref{yuk5}  depict the  new  $(\alpha,\, \gamma)$ intersection
areas in the third toroidal direction for the Up- and
Down-sectors, respectively. One can now explicitly calculate the
new areas by essentially employing the magnitude of  vectors
$(\vec {\cal A}_i)^2=(\epsilon_i {R^{(i)}_1})^2$ and $({\vec {\cal
B}_i})^2=(\epsilon_i {R^{(i)}_2})^2$   associated  with the sides
of the intersection triangles in the first two two-tori   for
 $U(1)_{8,8'}$ and $U(2)_L$   branes,  as well as  the corresponding vectors  ${\vec {\cal A}_{\alpha,\,\gamma}}$ and ${\vec {\cal
B}_{\alpha,\,\gamma}}$ in the  respective $U(1)_{8,8'}$ and
$U(2)_L$ sides of the intersection triangles 
of the third
two-torus. 
For the sake of simplicity  we set
$\epsilon_2=0$, since this significantly simplifies the analytic
expression for the intersection area, although the complete
formula is straightforward to obtain.  The  
intersection  area is:
\begin{eqnarray}
%
A^{U/D}_{\alpha, \, \gamma}
=\textstyle{1\over 2}\sqrt{ (\epsilon_1{R^{(1)}_1})^2\,
(\epsilon_1{R^{(1)}_2})^2 + (\epsilon_1 {R^{(1)}_1})^2\, {\vec
{\cal A}_{\alpha,\,\gamma}}^2+ (\epsilon_1 {R^{(1)}_2})^2\,
{\vec {\cal B}_{\alpha,\,\gamma}}^2 +4({\tilde
A}^{U/D}_{\alpha,\,\gamma})^2},\label{areaup}
\end{eqnarray}
 where ${\tilde A}^{U/D}_{\alpha,\,\gamma}$ refers to the corresponding
 intersection area in the
 third toroidal plane. The formula is valid as long as
$\epsilon_{1,3}$ are less or $\sim {1\over 2}$. For $U(1)_{8'}$
displacements the analogous
 area formulae are valid with the replacement $\epsilon_i\to\epsilon_i'$.



\medskip

Due to the orbifold and the orientifold symmetries, it is evident from
Figures \ref{yuk4} and \ref{yuk5} that a number of Up-Down Yukawa couplings
remain degenerate. In particular the following relations 
hold:
\begin{eqnarray}
A^{U}_{1,\,i}&=&A^D_{1,\,i} \nonumber\\
A^{U}_{3,\,ii}=A^D_{2,\,ii}&>&A^{U}_{2,\,ii}=A^D_{3,\,ii}\nonumber\\
A^{U}_{2,\,i}=A^D_{3,\,i}&>&A^{U}_{3,\,i}=A^D_{2,\,i}\nonumber\\
A^{U}_{1,\,ii}&<&A^D_{1,\,ii}  \label{areaud}
\end{eqnarray}
Except for the most suppressed Yukawa couplings between
the $\alpha=1$ Higgs
fields and $\gamma=ii$ quarks, the areas associated with the remaining
 Up- and Down-Yukawa couplings
pair-up.

 The explicit values for the areas and the vectors ${\vec{\cal
A}_{\alpha\, ,\gamma}}$
  and ${\vec{\cal B}_{\alpha\, ,\gamma}}$
 are given in Table \ref{up} for the Up-sector.
In the Up-sector  the area for $(3,i)$ is obtained from $(2,i)$ [and $(2,ii)$
from $(3,ii)$]  by changing
$\epsilon_3\to -\epsilon_3$. Similarly the Down-sector area for $(1,ii)$ is
obtained from the Up-sector area for $(1,ii)$ by  changing
$\epsilon_3\to -\epsilon_3$.


\begin{table}[htb] \footnotesize
\renewcommand{\arraystretch}{1.25}
\begin{center}
\begin{tabular}{|c|c|c|c|}
$(\alpha,\gamma)$   & ${\vec {\cal A}_{\alpha,\,\gamma}}^2$
 & ${\vec {\cal B}_{\alpha,\,\gamma}}^2$ \hspace{2cm}
& ${\tilde A}_{\alpha,\,\gamma}$ \hspace{1cm} \\
\hline $(1, i)$ & $\left(\textstyle{8\over
3}\epsilon_3R^{(3)}_1\right)^2$ & $\left(\textstyle{2\over
3}\epsilon_3R^{(3)}_1\right)^2+\left(\epsilon_3R^{(3)}_2\right)^2$
&
$\textstyle{4\over 3}\epsilon_3^2R^{(3)}_1R^{(3)}_2$ \\
\hline $(2,i)$    & $\left(\textstyle{4\over
3}(1+2\epsilon_3)R^{(3)}_1\right)^2$ & $\left(\textstyle{1\over
3}(1+2\epsilon_3)R^{(3)}_1\right)^2 +\left(\textstyle{1\over
2}(1+2\epsilon_3)R^{(3)}_2\right)^2$ &
$\textstyle{1\over 3}(1+2\epsilon_3)^2R^{(3)}_1R^{(3)}_2$ \\
\hline $(3,ii)$    & $\left(\textstyle{2\over
3}(1+4\epsilon_3)R^{(3)}_1\right)^2$ & $\left(\textstyle{1\over 6}
(1+4\epsilon_3)R^{(3)}_1\right)^2+\left(\textstyle{1\over
4}(1+4\epsilon_3)R^{(3)}_2\right)^2$ &
$\textstyle{1\over {12}}(1+4\epsilon_3)^2R^{(3)}_1R^{(3)}_2$ \\
\hline $(1,ii)$           & $\left(2(1-\textstyle{4\over
3}\epsilon_3)R^{(3)}_1\right)^2$ & $\left(\textstyle{1\over
2}(1-\textstyle{4\over 3}\epsilon_3)R^{(3)}_1\right)^2
+\left(\textstyle{3\over 4}(1-\textstyle{4\over 3}
\epsilon_3)R^{(3)}_2\right)^2$ & $\textstyle{3\over
4}(1-\textstyle{4\over 3}\epsilon_3)^2R^{(3)}_1R^{(3)}_2$
\end{tabular}
\end{center}
\caption{The values of ${\vec {\cal A}_{\alpha,\,\gamma}}^2$,
  ${\vec {\cal B}_{\alpha,\,\gamma}}^2$, and
${\tilde A}_{\alpha,\,\gamma}$ for various $\alpha$ and $\gamma$.
The results reduce to those in Table~\ref{veclep} for $\epsilon_3=0$.  }
\label{up}
\end{table}

\section{Implications of the Yukawa Coupling Hierarchy}
The basic results for the Yukawa couplings are given in equations
(\ref{Yukawa}), (\ref{area0}), (\ref{arealep}), and (\ref{areaup}).
It is difficult to discuss the implications for the fermion masses
without a detailed knowledge of the Higgs vacuum expectation
values (VEVs), which in turn depend on the details of the soft
supersymmetry breaking, the effective $\mu$ terms for the
Higgs fields, and the  normalization factors of the kinetic energy
terms, which have not been determined. 
As was discussed in~\cite{CLS}, the large number
of Higgs doublets and the lack of a compelling mechanism to generate
effective $\mu$ terms, at least at the perturbative level, are
significant drawbacks of the construction. Also, the construction
contains a strongly coupled quasi-hidden section, which is a 
candidate for dynamical supersymmetry breaking,  and the detailed
study of these phenomena is in progress \cite{clwf}.

Nevertheless, we can
make a few general comments about the implications of the
Yukawa couplings, emphasizing the simplest case in which the
D6 branes are positioned very close to the  symmetric positions in the
six-torus,
as in~(\ref{area0}). In this case, there are only four independent
Yukawa couplings, $h_{1,i}, h_{2,ii}= h_{3,ii},  h_{2,i}= h_{3,i}$,
and $h_{1,ii}$. As discussed in Section~\ref{yukawasection}
there are theoretical uncertainties concerning the prefactors
and numerical factors in the exponents. For definiteness,
we will assume
that $h_{\alpha,\gamma} \sim \exp(-A_{\alpha,\gamma}/(2\pi \alpha'))$
is a good approximation at least for the ratios of Yukawa couplings.
We will also assume that $M_s R_{1,2}^{(3)} \ge 2 \pi$. In that case,
$h_{1,i}\sim 1$, $h_{2,ii}= h_{3,ii} \le 0.59$,
$ h_{2,i}= h_{3,i} \le 0.12$, and $h_{1,ii} \le 0.009$, with
all but $h_{1,i}$ being extremely small for $ R_{1,2}^{(3)}$
much larger than the minimum value of $2 \pi/M_s$.
Intermediate values for the $ R_{1,2}^{(3)}$
will yield nontrivial hierarchies for the Yukawas.

It is convenient to rewrite (\ref{Yukawa}) as
\begin{eqnarray}
H_{yukawa}&=& Q_L^i \sum_{K=1}^4 \left[
h_{1,i} H^1_{UK} + \sqrt{2} h_{2,i} \left( \frac{H^2_{UK} + H^3_{UK}}{\sqrt{2}}
\right) \right] \bar{U}_K  \nonumber \\
&+&
Q_L^{ii} \sum_{K=1}^4 \left[
h_{1,ii} H^1_{UK} + \sqrt{2} h_{2,ii} \left( \frac{H^2_{UK} + H^3_{UK}}{\sqrt{2}}
\right) \right] \bar{U}_K, \label{Yukawa2} 
\end{eqnarray}
where the index $K$ represents the four terms ($I, II$, and the primed terms)
in~(\ref{Yukawa}).
When some of the Higgs fields acquire VEVs this will yield a $2 \times 4$
mass matrix for the two $U$ quarks and four antiquarks. However, in the special case
that the two rows are proportional (i.e., that they are aligned in the $K$
direction), there will only be a single nonzero mass eigenvalue.
Let us first consider the case of large sizes, so that all of the couplings
are small except $h_{1,i}$. Then, there will only be one significant
mass term, corresponding to $Q_L^i$ and a linear combination of the
$\bar{U}_K$, with coefficients depending on the VEVs of the $ H^1_{UK}$.
The other mass eigenvalue will be exponentially small.
In the special case of radiative symmetry breaking, usually associated with
supergravity mediated supersymmetry breaking but also
occurring for gauge mediation, the second mass would 
be exactly zero. That is because only the $ H^1_{UK}$'s have the
large Yukawa couplings needed to drive their (presumably positive)
mass-squares at the string scale to negative values at low energies,
and the VEVs of the other Higgs doublets would vanish. (The small
$h_{1,ii}$ would lead to a tiny mixing between $Q^1_L$ and $Q^2_L$,
but not generate a second non-zero mass because the two terms
would be aligned in $K$.) On the other hand, for small 
$R_{1,2}^{(3)}$ both $h_{1,i}$ and $\sqrt{2} h_{2,ii}$ 
(the Yukawa coupling for the relevant state $({H^2_{UK} + H^3_{UK}})/\sqrt{2}$),
could be significant, leading to two non-zero mass eigenstates provided
that the terms are not aligned in $K$.
For radiative breaking, the two large Yukawas could drive both relevant
mass-squares negative, and alignment would not be expected except for
very specific values for the mass-squares at the string scale and the effective
$\mu$ parameters and kinetic terms. 
In this case, the hierarchy $m_t \gg m_c$, $m_u =0$ could be achieved
by a hierarchy in the VEVs of the $ H^1_{UK}$'s relative to the
 $({H^2_{UK} + H^3_{UK}})/\sqrt{2}$), which could be achieved by modest
 differences in the relevant soft supersymmetry breaking and other terms.

Thus, it is possible to achieve a hierarchy of Up mass eigenvalues, associated
with the hierarchy of Yukawa couplings or of VEVs or both.
As discussed after  (\ref{areaud}), even after moving the branes
from their symmetric positions the Up and Down Yukawas are the
same up to relabelling except for the smallest coupling $h_{1,ii}$.
Thus, the hierarchy $m_t \gg m_b$ would have to come about
because the  $ H^1_{UK}$ VEVs are much larger than those for
the $ H^1_{DK}$, analogous to the large $\tan \beta$ region of the MSSM.
This can easily occur for moderate differences in the
soft mass-squares, especially if the effective $A$ parameters are small.
The full hierarchy of $m_t, m_b, m_c,$ and $m_s$ (with $m_d = m_u=0$)
could most likely be achieved for appropriate soft and effective $\mu$
parameters and kinetic energy terms, but we do not pursue this in
detail since these have not been calculated. Similarly, non-trivial
quark mixing could be generated by different $K$ dependence of the
VEVs in the $H_U$ and $H_D$ sectors.

In the symmetric case the charged lepton Yukawas
are the same as for the Up and Down quarks, and the charged leptons
couple to the same Higgs doublets as the Down quarks.
This is analogous to the $t-b-\tau$ Yukawa universality of the
simplest version of  $SO(10)$ grand unification, which
is  successful for large $\tan \beta$.
(In addition to the $b-\tau$ Yukawa relation at the string scale, one
also has $s - \mu$ unification. Of course, the quark Yukawas are
 enhanced by QCD and other effects in the running
 down from the string scale, leading to a successful $m_b/m_\tau$ prediction,
 but a rather large value for $m_s/m_\mu$.) The corrections to the lepton
 Yukawas from moving the $U(1)_1$ branes in (\ref{arealep}) decrease
 the lepton Yukawas relative to the symmetric quark couplings, 
 increasing the $m_b/m_\tau$ prediction. 
 Such a shift is acceptable as long as is small.
The $U(1)_{8,8'}$ shifts
 in (\ref{areaup}) have the same effects on the leptons as the quarks.

One expects Dirac neutrino masses comparable to the quark and
charged lepton masses close to the symmetric points. The possibility
of a neutrino seesaw was commented on in~\cite{CLS}. In particular,
Majorana masses for the right-handed neutrinos $\bar{N}$ cannot
be significantly larger than the scales at which the two additional
$U(1)'$ factors of the model are broken. It was shown that
when the charge confinement and anomaly conditions associated with
the strongly coupled quasi-hidden sector are taken into account,
then there would be scalar fields with the appropriate quantum
numbers to break both $U(1)'$s at a high scale at which the interactions
become strongly coupled. This could be $10^{15}$ GeV or higher,
which could lead to acceptable seesaw mass scales for the neutrinos.
However, the actual potential for those fields and their couplings
to the $\bar{N}$ states (needed to estimate the
actual masses and mixings) would be non-perturbative effects, beyond the
scope of this investigation.

\section{Conclusions}

We have considered the Yukawa couplings in a supersymmetric
three family Standard-like string Model. In particular, we have calculated
the leading order contributions to the world-sheet instantons associated
with the action of the string worldsheet stretching among the
intersection points corresponding to the chiral matter fields. 
We considered both the  case in which the branes are located
very close to symmetric positions in the six-torus, which leads to
a high degeneracy of Yukawa couplings, and the consequences of moving
some of the branes away from the symmetric positions. In general
there is a large hierarchy of Yukawa couplings, which increases
exponentially as the sizes of the tori are increased.
The actual fermion masses depend on the vacuum expectation values
of the Higgs fields, which in turn depend on the supersymmetry
breaking and on the effective $\mu$ parameters. There
are typically either two or one massive generations of
fermions.

\acknowledgments
We are especially grateful to Angel Uranga for many discussions and
collaboration at an early stage.
We would also like to thank Jing Wang
for useful discussions  and collaborations on related work.
This work was supported by
the DOE grants  EY-76-02-3071
and  DE-FG02-95ER40896;
by the National Science Foundation Grant No. PHY99-07949;
 by the University of Pennsylvania School of Arts and Sciences
Dean's fund (MC and GS) and Class of 1965 Endowed Term Chair (MC); by the
University of Wisconsin at Madison (PL);
by the W. M.
Keck Foundation as a Keck Distinguished Visiting Professor at the
Institute for Advanced Study (PL). We would  also like to thank
 ITP, Santa Barbara, during the ``Brane World'' workshop (MC, PL and GS), and
 the Isaac Newton Institute for Mathematical Sciences,
Cambridge, during the M-theory workshop (MC), for hospitality during the
course of the work.

\end{document}